\documentclass[aps,prb,twocolumn,floatfix,english,showpacs,10pt,superscriptaddress]{revtex4-2} 
\usepackage{amsmath,bm}
\usepackage{mathtools}
\usepackage{commath}
\usepackage{amssymb}
\usepackage{graphicx}
\usepackage{babel}
\usepackage{cases}
\usepackage[colorlinks=true, citecolor=blue, anchorcolor=green]{hyperref}
\hypersetup{linkcolor=red}
\usepackage{natbib}
\usepackage{floatrow}
\usepackage{dblfloatfix}
\usepackage{xcolor}
\usepackage{braket}
\usepackage{lipsum}

\makeatletter

\newcommand{\Rmnum}[1]{\expandafter\@slowromancap\romannumeral #1@}

\makeatother

\begin{document}
\title{Radiation-free and non-Hermitian topology inertial defect states of on-chip magnons}
\author{Bowen Zeng}
\affiliation{School of Physics and Electronics, Hunan University, Changsha 410082, China}
\author{Tao Yu}
\email{taoyuphy@hust.edu.cn}
\affiliation{School of Physics, Huazhong University of Science and Technology, Wuhan 430074, China}

\date{\today}

\begin{abstract}
Radiative damping is a strong dissipation source for the quantum emitters hybridized with propagating photons, electrons, or phonons, which is not easily avoidable for on-chip magnonic emitters as well that can radiate via the surface acoustic waves of the substrate. Here we demonstrate in an array of on-chip nano-magnets coupled in a long range via exchanging the surface acoustic waves that a point defect in the array, which can be introduced by the local magnon frequency shift by a local biased magnetic field or the absence of a magnetic wire, strongly localizes the magnons, in contrast to the well spreading  Bloch-like collective magnon modes in such an array setting. The radiation of the magnon defect states is exponentially suppressed by the distance of the defect to the array edges. Moreover, this defect state is strikingly inertial to the non-Hermitian topology that localizes all the extended states at one boundary. Such configuration robust magnon defect states towards radiation-free limit may be suitable for high-fidelity magnon quantum information storage in the future on-chip magnonic devices. 
\end{abstract}
\maketitle

\section{Introduction}

Defect bound states of electrons \cite{defect_state_electron}, phonons \cite{defect_state_phonon}, and photons \cite{defect_state_photon} are important factors affecting the optical and transport properties of solids, which may also act as candidates in the future quantum computing, such as the Majorana zero modes \cite{Majorana_1,Majorana_2} that may be treated as special defect state. These defect states are widely discovered in the Hermitian scenario when the dissipation plays a minor role. Verba \textit{et al.} predicted the magnon defect state in a two-dimensional array
of dipolarly coupled magnetic nanodots \cite{magnon_defect_state_2020}, acting as a Hermitian artificial system for reconfigurable signal processing and
microwave applications. 
Recent studies towards the non-trivial role of dissipation in the optic, phononic, electronic, and mechanic circuits open the routes to engineering the non-Hermitian topological states such as the exceptional nodal phase and non-Hermitian skin effect for the open system \cite{non_Hermitian_topology,WN_and_SE,Topoorigin}. In such a non-Hermitian scenario, the defect states are revealed to exist in non-Hermitian flatband of photonic zero modes \cite{defect_existence}, acquire the topological protection \cite{defect_topological_protection}, and possess flexible tunability by external sources \cite{defect_PT}. Magnons are the low-energy consumption information carriers \cite{magnonics_1,magnonics_2,magnonics_3,magnonics_4,magnonics_5}. 
In contrast to their electronic, acoustic, optic, and mechanic counterparts, the non-trivial role of dissipation is much less exploited in the magnonic circuits \cite{Flebus_perspective}. Recent studies proposed the realization of the non-Hermitian skin effect \cite{non_Hermitian_topology,WN_and_SE,Topoorigin} with much improved sensitivity in the magnetic nanowire array \cite{longrangeskin} and spin-orbit-coupled
van der Waals magnet \cite{SE2DM}.

Magnetic nanowire arrays---or the one-dimensional magnonic crystal---are excellent devices \cite{array_1,array_2,array_3,array_4} for microwave filtering \cite{filter}, magnetic recording \cite{recording}, and spin logics \cite{handbook}. Such on-chip magnonic devices are fabricated on the dielectric substrate, rendering that they may couple strongly to the surface acoustic waves of the proximity substrate \cite{magnon_SAW1,magnon_SAW2}. 
Indeed, recent studies demonstrated the magnetization dynamics of on-chip nano-magnets efficiently pumps phonons of the dielectric substrate via the magnetostriction \cite{GGGYIG,GGGmaterial,GGGattenuation,bright_dark_phonon,SAW_PRL,phononpumping1,phonontransport,magnetphonon,phonondiode}, achieving efficient transfer and communication of spin information over a long distance. Similar hybrid systems between the magnetic array and microwaves or superconducting circuits \cite{cavitymagnon,cavitymagnon2,magnonqubit} adds important opportunities to the non-reciprocal (quantum) information processing \cite{chiral_optics}. In such non-Hermitian systems, we reported that \textit{all} the bulk modes of magnons in a magnetic array are skewed at one edge by the non-Hermitian skin effect when the magnon interacts non-reciprocally or chirally in a finite range \cite{longrangeskin,Chirality_review}. However, the coupling between the on-chip magnons with the propagating waves brings an additional collective dissipation  \cite{magnonraditive,darkmode,magnongaccumulate,cavitymagnon,cavitymagnon2,magnonphoton,GGGYIG,SAW_PRL} with the damping rate probably much larger than the intrinsic one \cite{Gilbert}. Such a collective radiation is classified as subradiance and superradiance according to their suppressed and enhanced radiation rates to the individual one \cite{RMPlong, quantumRMP,mutiplescalingforsubradiant,Dicksuperradiant,zhangtheory,photonstorage,extremesub,emitter1,emitter2,emitter3}.  Recent experiment shows that the magnon dark modes preserving a long radiative lifetime induced by destructive interference between multiple magnetic spheres loaded in a microwave cavity can store the coherent information with magnons \cite{darkmode}.

\begin{figure}[htp]
\centering
\includegraphics[width=8.7cm]{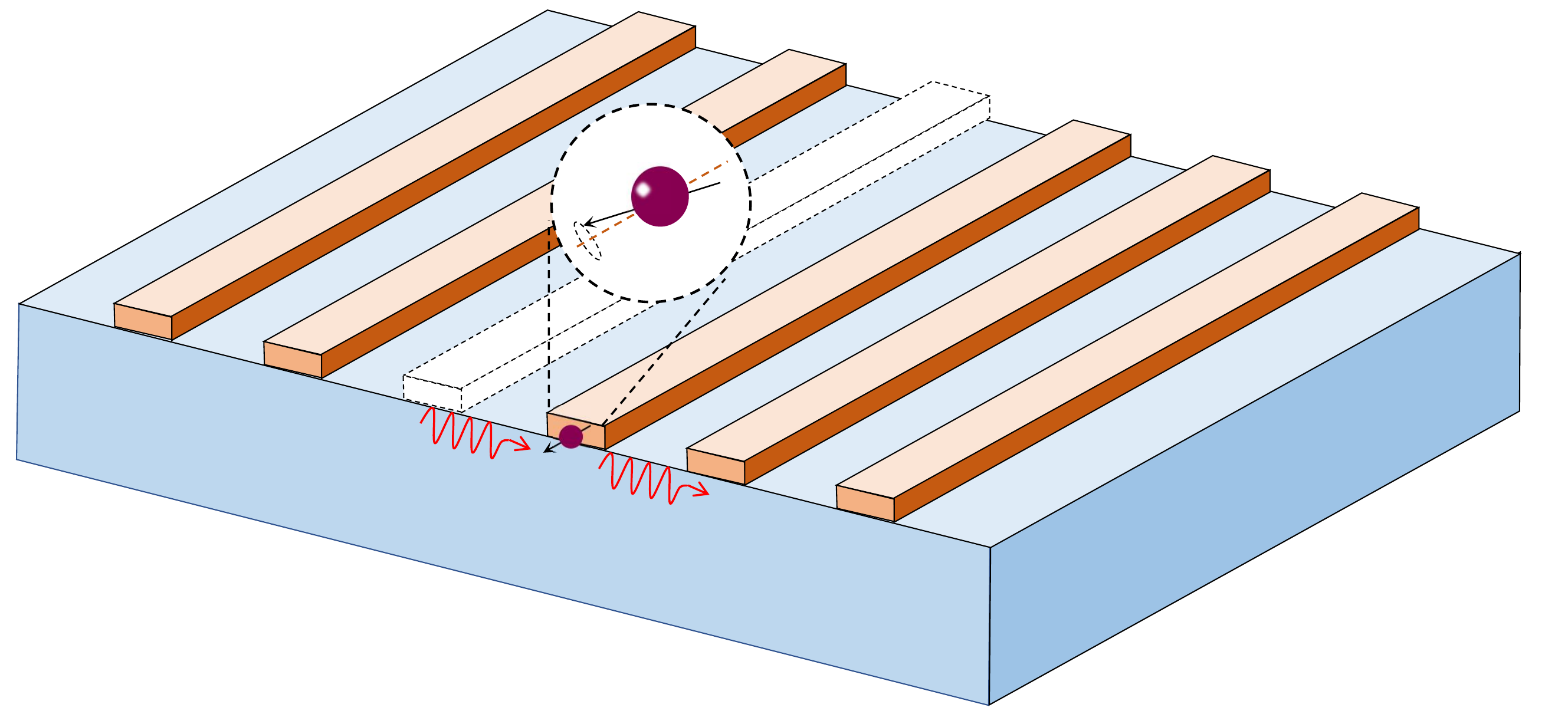}
\caption{Schematic of a magnetic nanowire array  fabricated on top of a dielectric substrate as one-dimensional magnonic crystal. A vacancy defect can be introduced by a local magnon frequency
shift by applying a local magnetic field or the absence of a magnetic nanowire at, e.g., the black dashed position. The uniform spin precession or Kittel mode in the magnetic wire is represented by the arrow precessing around the  axis fixed by the saturated magnetization, while the surface acoustic waves in the dielectric substrate are represented by the red waved lines. By virtually exciting and absorbing the proximity surface phonon via the magnetostriction, the magnons in different wires can indirectly interact with each other in a long range. }
\label{fig:model}
\end{figure}

In this work, we propose a scenario for realizing almost radiation-free modes of on-chip magnons in an array of magnetic wires \cite{array_1,array_2,array_3,array_4} that are fabricated on top of a thick dielectric substrate, as illustrated in Fig.~\ref{fig:model}, in which the magnons are dissipatively coupled in a long range via exchanging the surface acoustic waves of the substrate \cite{phonondiode}. We calculate the collective modes of the magnons when there exists a vacancy defect in the array, which can be introduced by the absence of a magnetic nanowire or a local magnon frequency
shift by applying a local biased magnetic field, and find that there is always a defect state, in spite of the presence of the long-range interaction, with which the spin fluctuation becomes strongly localized around the defect with a large amplitude that is even larger than those of all the other modes. Such features are preserved even when there exists the non-Hermitian skin effect with a special design \cite{longrangeskin}, a remarkable property that is inertial to the non-Hermitian topology when all the extended modes are skewed to one boundary, with which the defect state is well separated in space with all the other bulk modes. The lifetime of such defect state is much longer than all the other collective modes and is rarely affected by the defect position or the length of array, towards almost radiation-free limit when the array is long. We trace that the defect states occur via localization of the most subradiant states by a perturbation analysis via deforming the lattice constant of two neighboring wires. These findings may pave the way to potential application in such as high-fidelity information storage and single magnon trapping at defect in the quantum regime.

This paper is organised as follows. In Sec.~\ref{model} we first review the non-Hermitian effective Hamiltonian that describes the phonon-mediated long-range dissipative interaction between magnons in the magnonic array. Then we calculate the collective modes in the presence of the vacancy defect in the array, and address the  radiation property and non-Hermitian inertia  of the defect states in Sec.~\ref{robust}. To trace the localization mechanism, we perform a perturbation calculation in the non-Hermitian long-range coupled system in Sec.~\ref{perturbation_theory}. Finally, we discuss and summarize the results in Sec.~\ref{conclusion}.

\section{Dissipatively coupled magnons}

\label{model}

We consider here an array of $N$ on-chip magnetic nanowires \cite{array_1,array_2,array_3,array_4} with an equal spacing $d$ on top of a dielectric substrate, as shown in Fig.~\ref{fig:model}. These magnetic wires couple with the dielectric substrate via the magnetostriction. The wire distance $d$ is large such that may strongly suppress the dipolar interaction between wires \cite{magnon_defect_state_2020}, which is disregarded in this work. Such a hybridized system has been widely addressed recently in the context of phonon pumping by the magnetization dynamics \cite{GGGYIG,GGGmaterial,GGGattenuation,bright_dark_phonon,phononpumping1,phonontransport,magnetphonon,phonondiode}, among which a model Hamiltonian for the one-dimensional magnonic crystal is derived \cite{phonondiode,SAW_PRL,Chirality_review}. 
Below we review the key properties of such Hamiltonian.

Among the spin fluctuations in the magnetic nanowire, we focus on the Kittel magnon \cite{kittel}, which is quantized as $\hat{H}_m/\hbar=\omega_{m}\hat{m}^{\dagger}\hat{m}$. Here $\hat{m}$ is the magnon annihilation operator for the Kittel magnons with the frequency $\omega_{m}\rightarrow \mu_0 \gamma H_0$ under a strong magnetic field $H_0$ \cite{phonondiode}, with $\mu_0$ and $\gamma$ being the vacuum permeability and gyromagnetic ratio, respectively.
For the dielectric substrate, the acoustic modes that most strongly couple with the proximity magnets are the surface acoustic waves, which are
evanescent normal to the propagating plane \cite{Chirality_review,SAW_book}. They hold chirality since the rotation direction of their dynamical strain is locked to their propagation direction, namely, the generalized spin-momentum locking \cite{Chirality_review,SAW_book}. Due to its high sensitivity for the excitation and detection and low energy loss in the transmission, such surface acoustic waves are widely exploited in the signal detection, information transfer processing  \cite{phononpumping1,phonontransport,magnetphonon,phonondiode}, and manipulation of the magnetization dynamics \cite{magnon_SAW1,magnon_SAW2}.

As harmonic oscillators the surface acoustic waves with wave vector $k$ normal to the wire direction are  quantized as $\hat{H}_p/\hbar=\sum_k A_{k}\hat{a}_{k}^{\dagger}\hat{a}_{k}$,
in terms of the frequency $A_{k}=v |k|$, a constant group velocity $v$, and the phonon annihilation operator $\hat{a}_{k}$. The magnons $\hat{m}$ and surface phonons $\hat{a}_k$ are coupled via the magnetostriction, with the coupling strength $g_k$ being direction dependent and non-reciprocal with $|g_k|\neq |g_{-k}|$ in general by the chirality of surface acoustic waves and spin precession (for details we refer to Refs.~\cite{phonondiode,SAW_PRL}).  The total Hamiltonian reads
\cite{phonondiode}
\begin{align}\hat{H}/\hbar & =\sum_{j=1}^{N}\omega_{m}\hat{m}_{j}^{\dagger}\hat{m}_{j}+\sum_{k}A_{k}\hat{a}_{k}^{\dagger}\hat{a}_{k} \nonumber \\
 & +\sum_{j}\sum_{k}\left(g_{k} e^{ikz_{j}}\hat{m}_{j}\hat{a}_{k}^{\dagger}+{\rm H.c.}\right),
\end{align}
where $z_j=jd$ is the location of the $j$-th magnetic nanowire.

The magnon in a nanowire can virtually emit a surface phonon by the magnon-phonon coupling, and such phonon can propagate in a long distance before absorbed by the other nanowires, leading to a long-range coupling between magnons. When the Born-Markov
approximation works well, the coupled magnons in the $N$ nanowires mediated by the surface phonon may be described by a non-Hermitian Hamiltonian \cite{phonondiode,inoutput1,inoutput2,SAW_PRL}
\begin{align}
    \hat{H}_{\rm eff}/\hbar &= \left(\omega_m- i\frac{\Gamma _{R}+\Gamma _{L}}{2}\right) \sum_{j=1}^N \hat{m}_j^\dagger\hat{m}_j\nonumber \\
    & -i \Gamma_{L} \sum_{i<j} \beta_{k_0}^{|i-j|} \hat{m}_i^\dagger\hat{m}_j -i \Gamma_{R} \sum_{i>j} \beta_{k_0}^{|i-j|} \hat{m}_i^\dagger\hat{m}_j.
    \label{heff}
\end{align}  
Here, $\Gamma _R=\left|g_{k}\right|^2/v$ and $\Gamma _L=\left|g_{-k}\right|^2/v$ represent the emission rates of a single wire into the right and left propagating phonon modes. Since $|g_k|\ne |g_{-k}|$ by the chirality, $\Gamma_R\ne \Gamma_L$ in general. The exponential factor $\beta_{k_0}= \exp({ik_0d})$, in which $k_0$ denotes the resonant momentum of phonon. When disregarding the phonon attenuation, $k_0=\omega_m/v$ is real, but it acquires the imaginary components when the phonon has significant damping \cite{longrangeskin,phononattenuation}. Indeed, in high-equality elastic substrate such as gadolinium gallium garnet \cite{GGGYIG}, the imaginary part of $k_0$ can be safely disregarded \cite{GGGattenuation}. But there are cases that the imaginary part of $k_0$ has to be taken into account.  For example, considerable acoustic loss is reported in quantum state transfer using phonon \cite{phononattenuation}.

\section{Magnon defect states}
\label{robust}

We turn to address the nontrivial role of a point defect in the one-dimensional magnonic crystal on the collective modes of Hamiltonian (\ref{heff}).
The exact values of the parameters addressed above does not affect our findings to be addressed below, since the frequencies can be normalized by $(\Gamma_L+\Gamma_R)/2\sim O(1)~{\rm GHz}$ \cite{phonondiode}.
To be quantitative, we adopt $N=51$  yttrium iron garnet (YIG) nanowires \cite{YIG_wires} of equal neighboring distance $d=180$~nm with a fixed resonance frequency $\omega_m/2\pi=5.45 $~GHz \cite{phonontransport,bright_dark_phonon,GGGYIG}. The dielectric substrate with a negligible damping such as gadolinium gallium garnet (GGG) is of a choice, which is suitable for the long-range spin information transport \cite{phonontransport,bright_dark_phonon,GGGYIG}, with a large phonon propagating velocity  $v=3271$~m/s \cite{SAW1,SAW_book}. The resonant wavelength of surface acoustic waves to the ferromagnetic resonance is estimated to be $\lambda = 2\pi v/\omega_{m} \approx 600$~nm, with the corresponding wavevector $k_0=2 \pi/600$~nm$^{-1}$. With these parameters, we find the frequency and wavefunction of the magnonic collective modes via numerically solving the Hamiltonian (\ref{heff}).

\subsection{Frequency and wavefunction of collective modes}

\begin{figure}[ht]
\centering
\includegraphics[width=8.7cm]{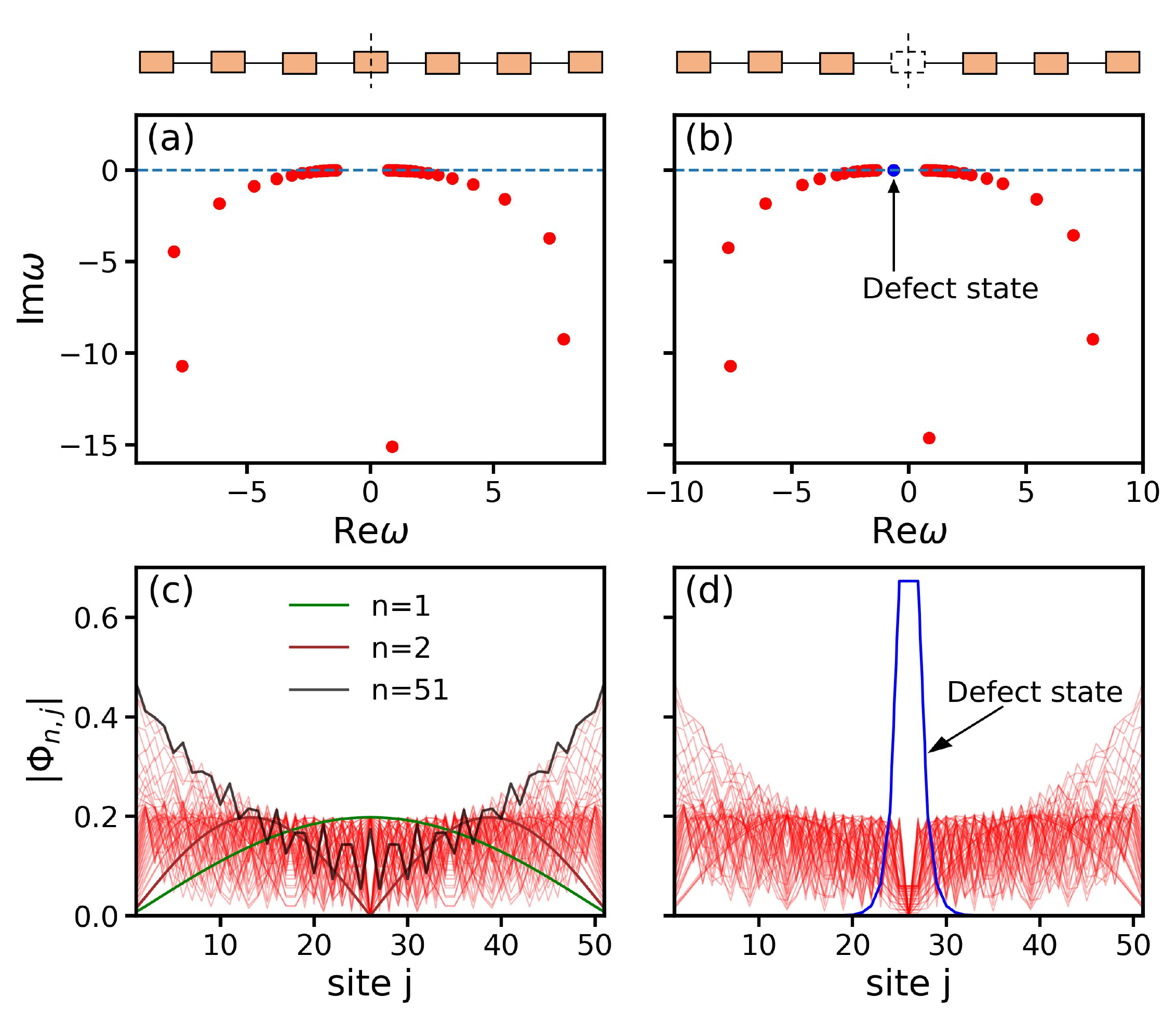}
\caption{Frequency spectra $\omega$ [(a, b)], scaled by $(\Gamma_R+\Gamma_L)/2$, and spatial profile $|\Phi_{n,j}|$ [(c, d)] of all the collective magnon modes without  [(a, c)] and with a point defect [(b, d)]. Here the radiation is non-chiral since $\Gamma_L=\Gamma_R$.  The side view of the magnetic wires is shown at the top of the figures.
The frequency of the defect states is emphasized by the blue dot in (b), with mode profile around the defect at the middle of the array ($j=26$) plotted in (d). In (c), we highlight two typical subradiant states ($n=1,2$) and sueprradiant state ($n=51$). }
\label{fig2}
\end{figure}

We now compare the frequency spectra $\omega$ and spatial profile (wavefunction) of the collective magnon modes $|\Phi_{n,j}|$ with and without a point defect in the array, where $n=\{1,\cdots,N\}$ are the mode indexes according to the increased decay rates, as shown in Fig.~\ref{fig2}.  The side view of the magnetic nanowire array is plotted at the top of the figures, where each rectangular block represents a nanowire and the defect is implied by a dashed box. As a frequency reference, the magnon bare frequency $\omega_m$ is dropped when plotting the frequency spectra, and the frequency is normalized by $(\Gamma_R+\Gamma_L)/2$. In this part, the radiation is assumed to be non-chiral with $\Gamma_L=\Gamma_R$, but the chirality plays an important role that will be addressed later. The wavefunction we plot is the right eigenvector $\Phi_{n}$ of the Hamiltonian (\ref{heff}), for which we note the left eigenvector $\Psi_{n}$ is no longer the Hermitian conjugation of the right eigenvector since the Hamiltonian is not Hermitian. The eigenvectors are normalized according to $\Psi_{n}^{\dagger}\Phi_{n}=1$.

Figure~\ref{fig2}(a) plots the frequencies of all the collective magnon modes for an array without any defect in terms of the real and imaginary components, where most states possess decay rates smaller than $(\Gamma_L+\Gamma_R)/2$, while there are several modes having decay rates much larger than $(\Gamma_R+\Gamma_L)/2$. The former are known as the subradiant states, while the latter are the superradiant states \cite{RMPlong, quantumRMP,mutiplescalingforsubradiant,Dicksuperradiant,zhangtheory,photonstorage,extremesub,emitter1,emitter2,emitter3}. We sort these frequencies according to the increased decay rates in terms of integers $n\in \{1,2,...,51\}$, such that $n=1$ and $51$ represent the most subradiant state and most superradiant state, respectively. 
The spatial profile of typical subradiant states ($n=1, 2$) and superradiant state ($n=51$) are plotted in Fig.~\ref{fig2}(c). We observe that these two subradiant states are extended over the bulk forming the standing waves with small amplitudes at the edge, while the superradiant state has a large amplitude at the edge. Such a weak skin tendency may be understood from a tight-binding Hamiltonian after taking the inverse matrix of the Hamiltonian \eqref{heff} that involves only the nearest hopping but additional local potential at the edge \cite{inverse}.

Now we introduce a defect in the middle ($j=26$) of the array. It is noticed that  a special frequency is then induced by the defect in the frequency spectra denoted by the blue dot in Fig.~\ref{fig2}(b), which does not present in Fig.~\ref{fig2}(a) and is isolated from the other frequencies with decay rate even smaller than the most subradiant state (more comparison to be addressed in Sec.~\ref{Radiation_rate} below). 
Figure~\ref{fig2}(d) shows the spatial profile of the defect state denoted by the blue color, which has an impressively larger amplitude than those of the other collective modes.  Such a defect state is strongly localized around the defect with a narrow spreading. On the other hand, the spatial profile of the other states are rarely affected, except for a strong attenuation around the defect position, a feature that is related to the long-range nature of the magnon coupling.

Recent experiment \cite{array_trapping} finds a strong trapping of magnons that are launched by the ferromagnetic resonance with a considerable amplitude at particular positions in a one-dimensional magnetic array. Such a observation appears to support our prediction of the magnon defect states.

\subsection{Radiation property of defect state}

\label{Radiation_rate}

Previous studies stated that the main dissipative channel for the collective modes governed by the Hamiltonian (\ref{heff}) is the radiation to the environment through the edge \cite{mutiplescalingforsubradiant}. Such an edge leakage is indeed strong for the superradiant states because they have larger amplitudes at the edge.  Thereby the strong localization of the defect state could efficiently prevent the radiation of surface acoustic waves through the edge, thus favoring an almost radiation-free mode in the open system. Such a mechanism is distinct from that of the bound state in the continuum \cite{BIC_1,BIC_2}, although similar excellent property can be achieved.  
We substantiate such expectation by analyzing the radiation property of the defect state by changing the position of the defect in the array, the length of the array,  and the chirality of the coupling with $\Gamma_L\ne \Gamma_R$.

\begin{figure}[ht]
\centering
\includegraphics[width=8.72cm]{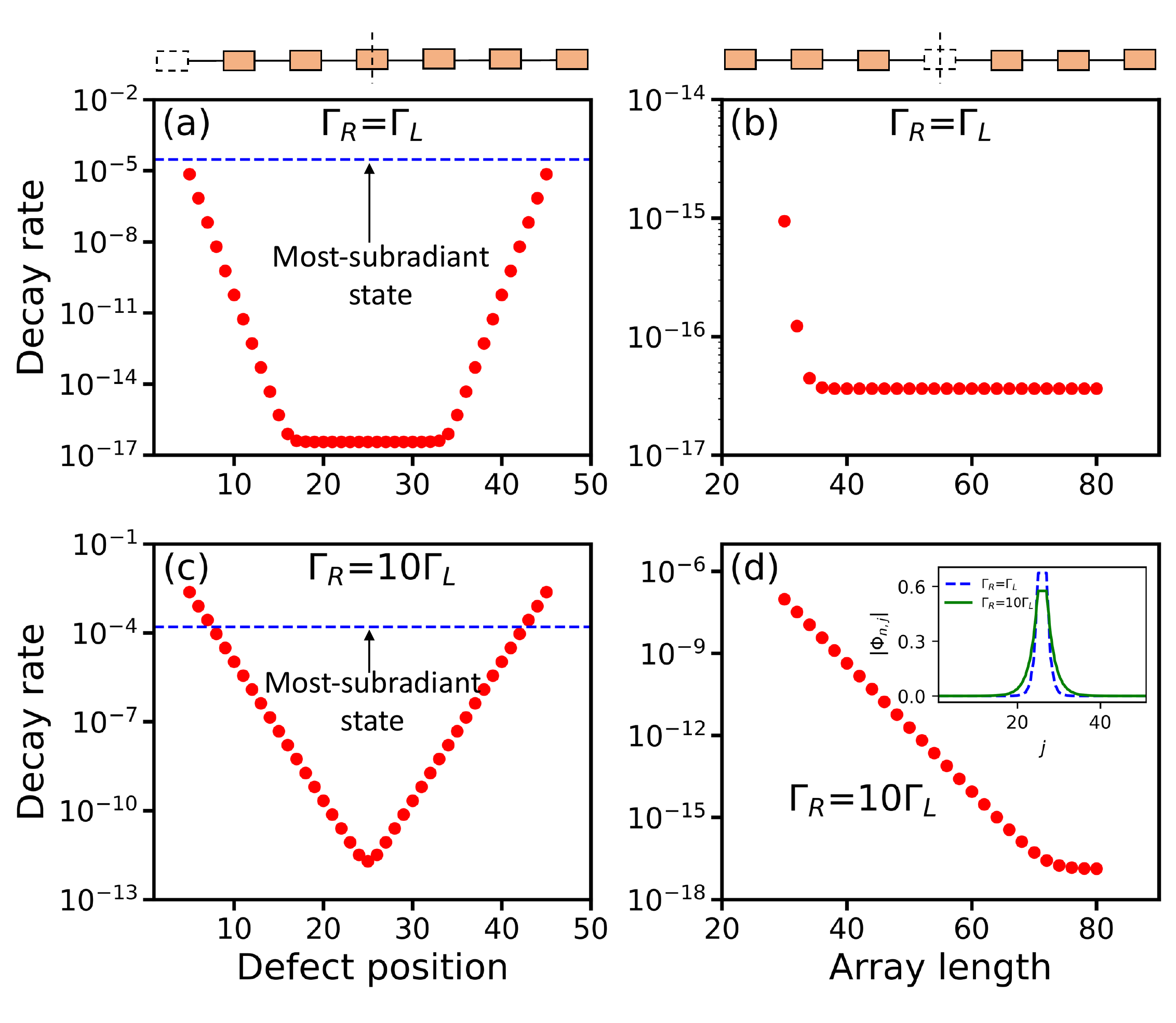}
\caption{Radiation properties of the magnon defect states revealed by the dependence of the radiation rates  on the defect position, array length, and coupling chirality. The side view of the magnetic wires is shown at the top of the figures, with the defect implied by the dashed rectangular box. (a) and (c) address the strong dependence of the decay rates of the defect states on the position in the array without ($\Gamma_R=\Gamma_L$) and with ($\Gamma_R=10 \Gamma_L$) the chirality. The decay rate of the most subradiant state is implied by the blue dashed line. In (b) and (d), the array length is changed to show the effect of spatial spreading of the defect states on the radiation.  We compare the spatial profile without and with the chirality in the inset of (d).} 
\label{fig3}
\end{figure}

 Since we expect that the more localized of the defect states away from the edge, the smaller of their decay rates,  we place the defect at different positions in the array and calculate their decay rates. Indeed, as plotted in Fig.~\ref{fig3}(a), the decay rates of the defect states change dramatically by many orders when located from the edge to the middle of the array. These decay rates are all smaller than that of the most subradiant state that is indicated by the blue dashed line in Fig.~\ref{fig3}(a), but this property is not always robust when the chirality kicks in as addressed in Fig.~\ref{fig3}(c). Crucially, the decay rates become negligibly small when the defect is located far away from the two edges, i.e., around the middle of the array. Therefore, the narrow spreading of the spatial  distribution leads to a negligible amplitude at the edge in a long array that strongly suppresses the radiation to the environment. This feature is further verified by changing the length of the array. Here we focus on the defect state at the middle of the array and plot in Fig.~\ref{fig3}(b) that by changing the length of array, the decay rates of the defect are unchanged in a long array but are strongly enhanced when the array is short. So it is clear that as long as the array length is much longer than the spatial spreading of the defect state, its decay rate is always almost negligible, a feature that is essential for high-fidelity information storage with on-chip magnons.

Chirality does affect the properties of the radiation, but the essential features of the radiation in the absence of the chirality are retained as shown in Fig.~\ref{fig3}(c) and (d) with $\Gamma_R=10\Gamma_L$. The radiation is still most suppressed when the defect is at the middle of the array as addressed in Fig.~\ref{fig3}(c), so without asymmetric property although the radiation is directional flowing to the right. A key 
feature is that the spatial profile of the defect state has a wider spreading than that without chirality, as shown by the inset of Fig.~\ref{fig3}(d), which leads to a stronger radiation. Thereby, achieving the same low radiation rate may need a longer array as in Fig.~\ref{fig3}(d).

We note that our conclusions with magnonic emitters apply to the quantum dipolar emitters as well such as the cold atoms coupled to the waveguide \cite{zhangscaling,coldatom,emitter1}, where the long-lifetime excitation for quantum processing remains wanting. It appears that both the defect and subradiant states have a long lifetime that may facilitate the  high-fidelity information storage. Here we highlight the merit of the defect state in two aspects. On the one hand, the subradiant state arises from the destructive interference and thereby may be more sensitive to the disorders such as the position fluctuation of the wires and the bare frequency broadening among the magnets.  While the defect state is well localized that may be difficult to be affected since it has minor overlapping with the other states [refer to Fig.~\ref{fig2}(d)]. On the other hand, different from the subradiant state, of which the decay rate obeys a scaling law as $N^{-3}$ \cite{zhangscaling,yuchiralmagnon,yuchiralmagnon2}, the defect state is independent of the length of a long array. For the subradiant state, it takes $N \approx 10^4$ to reach the same low decay rate as that of the defect state, which is of course far beyond the experimental capability with magnons or other quantum emitters such as cold atoms \cite{zhangscaling,coldatom,emitter1,emitter2,emitter3}.  Therefore, such a magnon defect state provides a robust and experimentally feasible platform to realize an almost radiation-free state even with not too many magnetic wires, say $N \lesssim 20$.

\subsection{Inertia to non-Hermitian skin effect}

According to our previous work \cite{longrangeskin}, the subradiant and superradiant magnon states \textit{all} become asymmetrically skewed to one boundary of the array when the non-Hermitian skin effect occurs. Such an effect appears as a non-Hermitian topological phenomena \cite{non_Hermitian_topology,WN_and_SE} when the interaction between the magnons is chiral and the phonon is attenuated in the propagation, rendering a relatively short-range non-reciprocal interaction. Since the topological effect is robust to the disorder, one could conclude that the defect state had been also affected and skewed to one boundary by the non-Hermitian skin effect. However, we find that the defect state is very inertial to such a topological phenomena, thus demonstrating the superior advantage of such a defect localization for memory functionality.

\begin{figure}[ht]
\centering
\includegraphics[width=8.7cm]{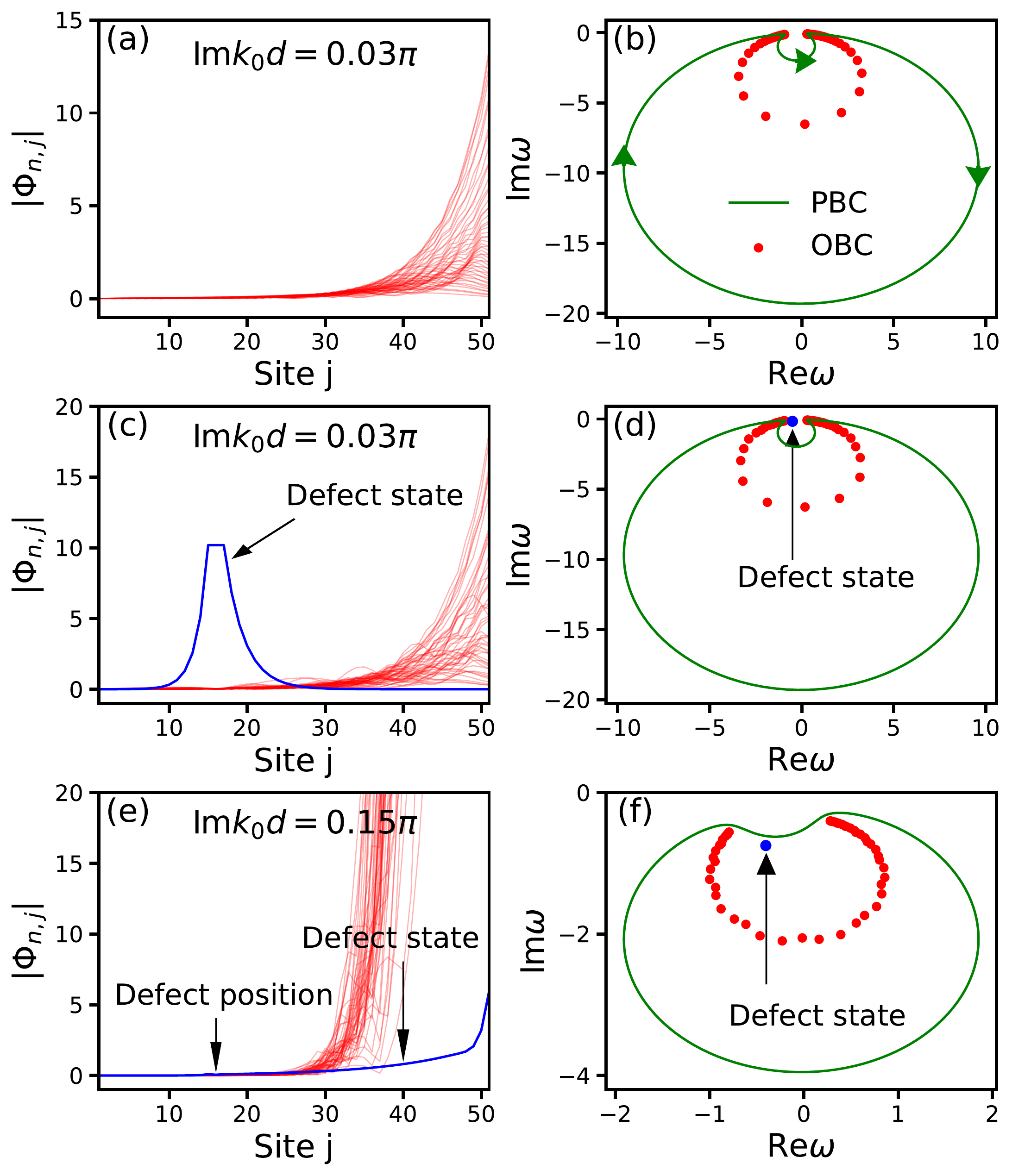}
\caption{Inertia of the magnon defect state to the non-Hermitian skin effect under the chiral coupling $\Gamma_R=10\Gamma_L$ and phonon attenuation. (a) shows the skewed localization of all the magnon states under the phonon attenuation $\text{Im}k_0 d=0.03\pi$ in the absence of a defect, with their frequency distribution under OBC (red dots) and PBC (green curve) addressed in (b). (c) shows a spatial well isolated defect state by the blue curve when introducing a defect. The frequency of such a defect state is not enclosed by the frequency winding under PBC, as spotted in (d) by the blue dot. However, when the phonon attenuation is very strong with $\text{Im}k_0 d=0.15\pi$ that favors a short-range chiral coupling between magnons, the defect state is skewed with its frequency well enclosed by the frequency winding, as shown in (e) and (f), respectively.}
\label{fig4}
\end{figure}

In Fig.~\ref{fig4}(a), we show the asymmetrically skewed localization of all the magnon states at the right edge of the array without defects when the phonon mediated interaction is chiral $\Gamma_R=10\Gamma_L$ and the phonon is attenuated during propagation with $\text{Im}k_0 d=0.03\pi$ that corresponds to the phonon attenuation length $200d/3=12~{\mu}$m, demonstrating again the non-Hermitian skin effect \cite{longrangeskin}. We note that such a phonon attenuation length is longer than the array length $50d$, so the non-reciprocal effective interaction is still of quite long range. The topological origin of the skin effect is addressed in Fig.~\ref{fig4}(b), in which we plot the normalized frequencies under the open boundary condition (OBC) by the red dots and under the periodic boundary condition (PBC) by the blue curve, both calculated from Eq.~(\ref{energy}) in the Appendix~\ref{appendix_A}. With PBC, the Bloch vector $\in[-\pi/d,\pi/d]$ is real. When it evolves from $-\pi/d$ to $\pi/d$, the normalized frequency evolves clock-wisely as indicated by the arrow. The corresponding topology is captured by the so-called winding number \cite{toposkin,windingnumber2,mutiplewinding} that counts the times of the spectra under PBC encloses a reference energy, which is minus unity in this case (refer to Appendix~\ref{appendix_A} for the details), so being topologically nontrivial.

Now we introduce a defect at the left half of the array ($j=16$), a design avoiding the interference with the skin modes that are localized at the right boundary.  The distribution of all the states with such a point defect is plotted in Fig.~\ref{fig4}(c). Interestingly, the defect state survives with a considerable amplitude around the defect position, indicating that the defect state is insensitive to the non-Hermitian skin effect in such a long-range coupled system. While the other states are all skewed to the right boundary by the skin effect, which achieves a spatial separation between the defect state and all the other collective modes. 
This feature is unique since it distinguishes from the defect state in the Hermitian counterpart, where there is always a spatial overlap with the bulk states. Such a spatially well isolated magnon defect state may find some important application in on-chip quantum information storage such as trapping, local detection, and local manipulation of single magnon state \cite{quantum_magnonics}.

For the inertia of the defect state to the non-Hermitian topology we trace a close relation between the frequency winding under the PBC, i.e., those green circles plotted in Fig.~\ref{fig4}(b), (d), and (f), and the location of the frequency of the defect state. Such inertia appears to depend on whether the defect frequency lies in the winding circle or not. The frequency of the defect state in Fig.~\ref{fig4}(c) indeed locates outside of the winding circle, as plotted in Fig.~\ref{fig4}(d), so being not affected by the non-Hermitian topology. As a comparison, we now change the parameters arbitrarily with a very strong phonon attenuation $\text{Im}k_0 d=0.15\pi$ that corresponds to the attenuation length $2.4~{\rm \mu}$m of the surface acoustic waves. The defect state is completely skewed to the right edge as in Fig.~\ref{fig4}(e) and at the meantime, the defect frequency is located inside the winding circle in Fig.~\ref{fig4}(f). This highlights again the nontrivial role of the long-range coupling mediated by phonons in a high-quality substrate on the collective modes that renders robust trapping properties possible that is absent in the  Hatano-Nelson model with a short-range asymmetric coupling \cite{Hatano_Nelson}.

\section{Localization Mechanism}
\label{perturbation_theory}

So far we have addressed the emergence of a magnon defect state in a long-range coupled array and its properties in the radiation and robustness, but have postponed the discussion of its mechanism. Here we address this issue via a perturbation analysis on the effect of interference on the localization when locally deforming the lattice constant of two middle neighboring wires in the array. We find that the most subradiant state can evolve into a  localized state due to its interference with the other subradiant states, which further suppresses the radiation towards the almost radiation-free limit.

\begin{figure}[ht]
\centering
\includegraphics[width=8.8cm]{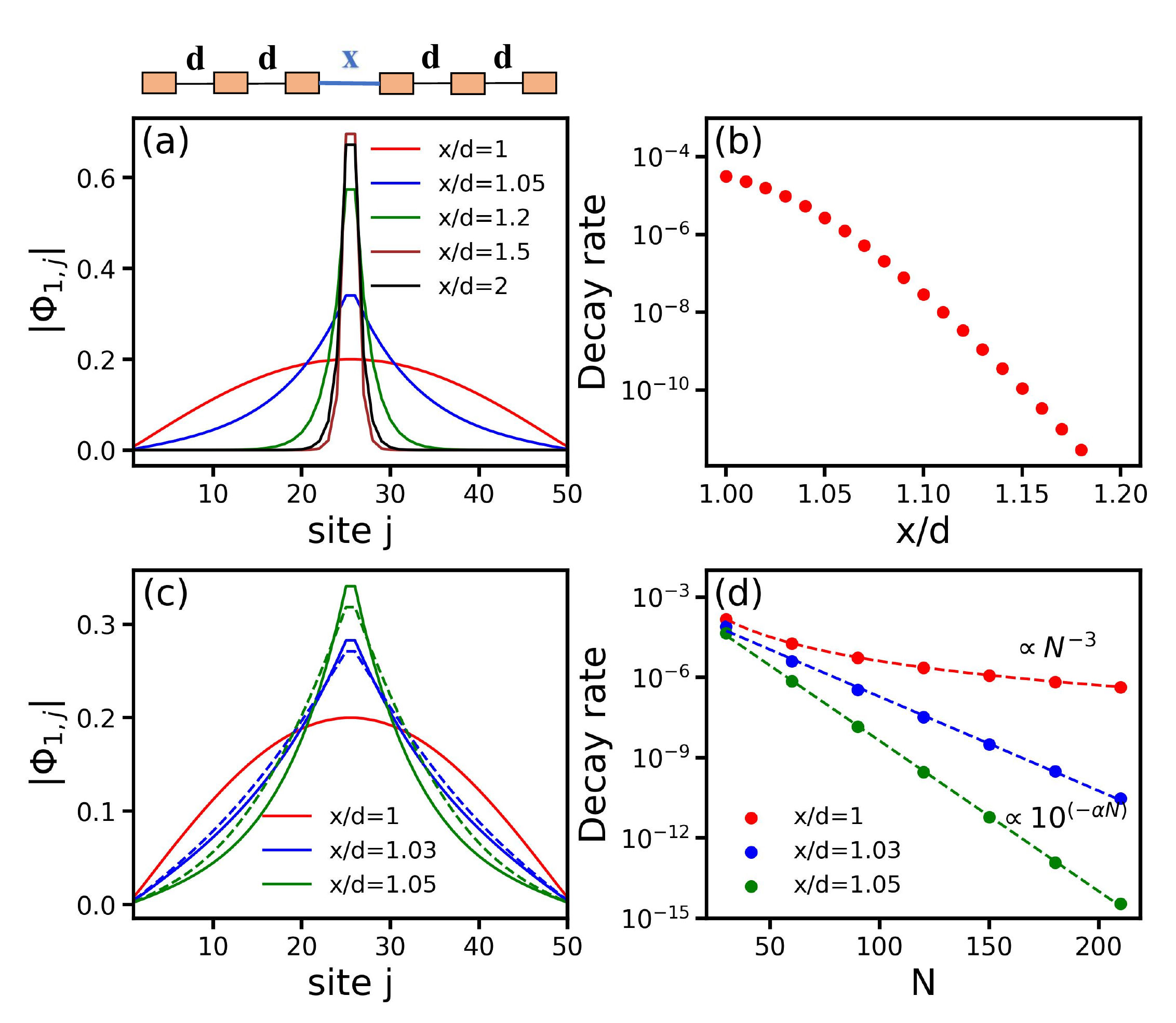}
\caption{Localization of the most subradiant state by the local change of the distance $x$ of two neighbouring wires. The configuration is illustrated at the top of (a). (a) shows the gradual localization of the  state with
longest lifetime when the spacing $x$ increases from $d$ to $2d$, with their decay rates shown in (b).
(c) compares the spatial profile of such states from the exact matrix diagonalization (the solid curves) and from the perturbation analysis (the dashed curves).  The dependence of the decay rates of these states on the array length is shown in (d), which follows exponential $N^{-3}$ when $x = d$ and a power law governed by
$10^{-\alpha N}$ with a constant $\alpha$ when $x>d$.}
\label{fig5}
\end{figure}

Intuitively, we may consider in another point of view the effect of a defect on the collective radiation.  Introducing a point defect in the array is equivalent to a local change of the distance of two neighboring wires from $x=d$ to $2d$ among the other, as illustrated at the top of Fig.~\ref{fig5}(a). Such a procedure can be achieved gradually via changing the distance $x$ from $d$ to $2d$, such that we can trace the evolution of the states, e.g., how one can arrive at a state with longer lifetime and strong localization from those states with shorter lifetime and wider spatial spreading.
Indeed, as shown in Fig.~\ref{fig5}(a), when the spacing $x$ increases, the spatial distribution of the state with longest lifetime becomes gradually localized at the center. Such a localization leads to a reduction of the decay rate in several orders of magnitude, as plotted in Fig.~\ref{fig5}(b), in consistent with the previous analysis in Sec.~\ref{robust}.  The continuous changes of spatial profile and decay rate suggest that a defect state is possibly evolved from the most subradiant state, as implied by a perturbation analysis with $x=d+\delta $ when $\delta\ll d$ below.

We express Eq.~(\ref{heff}) in the matrix form
$ \hat{H}_{\rm eff}/\hbar=\hat{M}^{\dagger}{\cal H}_{M} \hat{M}$,
where $\hat{M}=\left(\hat{m}_1,\hat{m}_2,\cdots,\hat{m}_N\right)^T$. The non-Hermitian matrix ${\cal H}_{M}$ can be diagonalized with biorthonormal basis $\left\langle \Psi_{l}\right|{\cal H}_{M}\left|\Phi_{n}\right\rangle  =E_{n}\delta_{ln}$ with $\Psi_{l}$ being the $l$-th left eigenvector and $\Phi_{n}$ being the $n$-th right eigenvector \cite{Biorthogonal}.
By a small distortion $\delta$, the total Hamiltonian ${\cal H}_{\rm M}\rightarrow {\cal H}_{\rm M}+\left(e^{ik_{0}\delta}-1\right){\cal H}'$, where the matrix is of the form
\begin{align}
{\cal H}'=\left(\begin{array}{cccccc}
... & ... & ... & ... & ... & ...\\
... & 0 & 0 & -ie^{2ik_{0}d} & -ie^{3ik_{0}d} & ...\\
... & 0 & 0 & -ie^{ik_{0}d} & -ie^{2ik_{0}d} & ...\\
... & -ie^{2ik_{0}d} & -ie^{ik_{0}d} & 0 & 0 & ...\\
... & -ie^{3ik_{0}d} & -ie^{2ik_{0}d} & 0 & 0 & ...\\
... & ... & ... & ... & ... & ...
\end{array}\right).
\end{align}
As $\delta$ is small, $e^{ik_{0}\delta}-1\approx ik_{0} \delta$, so the term ${\cal H}'$ is a perturbation, with which we can find the wavefunction by the first-order correction in the perturbation theory, which for the $n$-th right eigenvector is 
\begin{equation}\label{perturbation}
   \left| \Phi_{n}^{(1)} \right\rangle =  \sum_{l\neq n}\frac{\left\langle \Psi_{l} \right|\left(e^{ik_0\delta}-1\right){\cal H}' \left| \Phi_{n} \right\rangle}{E_n-E_l}\left| \Phi_{l} \right\rangle.
\end{equation}

We compare in Fig.~\ref{fig5}(c) the spatial profile of the state with the longest lifetime from the exact matrix diagonalization (the solid curves) and from the above perturbation analysis (the dashed curves). We observe that the  wavefunction from the perturbation analysis agrees well with the exactly calculated wavefunction when $x$ varies from $1.01d$ to $1.05d$, and captures the localization tendency by the ``defect", indicating that Eq.~\eqref{perturbation} is an excellent approximation. In Eq.~\eqref{perturbation}, the localized state is mainly contributed by the most subradiant state that superposes with the other subradiant states, indicating that the interference between them can lead to a more localized spatial profile. Besides, the interference also suppresses the decay rate as illustrated in Fig.~\ref{fig5}(d), in which we perform the calculation with different length of the array. Furthermore, the decay rate of the state with the longest lifetime follows a universal scaling law with exponential $N^{-3}$ when $x=d$, in consistent with the previous studies \cite{zhangscaling,yuchiralmagnon,yuchiralmagnon2}, but becomes a power law governed by $10^{-\alpha N}$ with a constant $\alpha$ when $x>d$.
This implies that the radiation of the magnon defect state is exponentially
suppressed by the distance of the defect to the array edge, a  property superior over that of the most subradiant states.

Recent studies reported analogous exponentially suppressed decay rate in the coupled atomic array mediated by light, such as assembling atoms in a ring waveguide to eliminate the boundary dissipation or applying  local deformation for the lattice constant to induce localized resonance \cite{mutiplescalingforsubradiant}. The on-chip magnons acts as magnonic quantum emitters that performs at the room temperature, which may help circumventing the harsh experimental environment such as fine control and extremely low temperature  required for the cold atom system.

\section{Discussion and conclusion}
\label{conclusion}

Defect is an important design parameter that may strongly affect the device performance, but it is difficult to be avoided in the fabrication of on-chip magnonic devices such as the magnetic nanowire arrays addressed here \cite{array_1,array_2,array_3,array_4}. On the other hand, the defect can also be specially designed by, such as a local magnon frequency shift by applying a local biased magnetic field, to achieve localized modes with advanced functionality such as the enhancement of the interaction between the magnonic quantum emitters with the other degrees of freedom \cite{optomagnon,influenceofdefect}. It might be surprising that the defect state can survive in the long-range coupled system and under the non-Hermtian skin effect, since both leads to strong delocalization effect.
The strong localization of magnons can be detected by the NV center magnetometry \cite{NV} coherently and the Brillouin light scattering incoherently \cite{BLS}.

In conclusion, we have demonstrated a strongly localized and almost radiation-free magnonic defect state introduced by a point defect in a on-chip  magnetic array that is coupled in a long range mediated by the surface acoustic waves of the substrate. Such a defect state is demonstrated to be even inertial to the non-Hermitian topology, protected by the long-range nature of the phonon-mediated interaction, although all the other states are skewed to one boundary. We find that the local deformation of the lattice constant induces the interference of the subradiant magnonic states, i.e., those collective magnon modes with longer lifetime than that of the individual one, which is responsible for the localization and a much longer lifetime.
The radiation-free and configuration robust defect state may be desired for many practical applications, such as the high-fidelity information storage and single magnon trapping. Our formalism on the magnonic quantum emitters can be extended into the other quantum dipolar emitters \cite{RMPlong,mutiplescalingforsubradiant,zhangscaling, zhangtheory,coldatom,emitter1,emitter2,emitter3}, and opens new perspective on the realization ultra-long lifetime states for quantum memory.

\vskip0.25cm	
\begin{acknowledgments}
We gratefully acknowledge Yu-Xiang Zhang, Jian-Song Pan, and Ke-Qiu Chen for many useful discussions. This work is financially supported by the startup grant of Huazhong University of Science and Technology (Grants No. 3004012185 and No. 3004012198) as well as the National Natural Science Foundation of China Grant No. 12004106.
\end{acknowledgments}

\begin{appendix}

\section{Analysis of non-Hermitian skin effect}

\label{appendix_A}

Here we derive the frequency spectra and the wavefunction  for the non-Hermitian Hamiltonian Eq.~\eqref{heff} and address the condition for the non-Hermitian skin effect and its topological origin under the long-range coupling. For the Hamiltonian $\hat{H}_{\rm eff}=\hat{M}^{\dagger}{\cal H}_{M} \hat{M}$ \eqref{heff}, we construct the trial solution $\hat{\varXi}_{\kappa}^{\dagger}=\hat{M}^{\dagger}\Phi_{\kappa}$ with the Bloch basis 
$\Phi_{\kappa}=(\beta_{\kappa}^1,\beta_{\kappa}^2, \cdots, \beta_{\kappa}^{N})^T$, following our previous works \cite{longrangeskin,magnongaccumulate}, where $\beta _{\kappa} \equiv \exp({i \kappa d})$ is expressed by the to be found complex wavevector $\kappa$.  
The equation of motion of $ \hat{\varXi}_{\kappa}^{\dagger}$ is governed by
\begin{equation}
   i {d \hat{\varXi}_{\kappa}^{\dagger}}/{dt}=-\omega_{\kappa} \hat{\varXi}_{\kappa}^{\dagger}+\Gamma_{R}g_{\kappa} \hat{\varXi}_{k_0}-\Gamma_{L}h_{\kappa} \hat{\varXi}_{-k_0},
\end{equation}
in which the complex dispersion relation \begin{equation}
\omega_{\kappa}=\frac{\Gamma_{L}}{2}\frac{\beta_{-\kappa}+\beta_{k_0}}{\beta_{-\kappa}-\beta_{k_0}}+\frac{\Gamma_{R}}{2}\frac{\beta_{\kappa}+\beta_{k_0}}{\beta_{\kappa}-\beta_{k_0}}
\label{energy}
\end{equation}
has two roots, say $\{\beta_{\kappa_1},\beta_{\kappa_2}\}$,
and $g_{\kappa}={\beta_{\kappa}}/({\beta_{\kappa}-\beta_{k_0}})$ and $h_{\kappa}={\left(\beta_{\kappa}\beta_{k_0}\right)^{N+1}}/({\beta_{\kappa}\beta_{k_0}-1})$ are dimensionless that modulates the coupling  constant of the modes with (complex) momentum $\pm k_0$ to the phonon. The residue coupling of $\hat{\Xi}_{\pm k_0}$ to the phonon is thereby sensitive to the chirality $\Gamma_L\ne \Gamma_R$.
\begin{align}
    \omega_{\kappa_1}=\omega_{\kappa_2}
    \label{condition_1}
\end{align} becomes the desired eigenvalue when the two roots satisfy 
\begin{equation}
g_{\kappa_2}h_{\kappa_1}=g_{\kappa_1}h_{\kappa_2},
\label{condition_2}
\end{equation}
and thereby the eigenstate of the problem is simply a superposition of two modes that does not have a definite momentum
\begin{equation}
\Phi=\frac{ \beta_{\kappa_2}}{ \beta_{\kappa_2}-\beta_{k_0}}\Phi_{\kappa_1} -\frac{ \beta_{\kappa_1}}{ \beta_{\kappa_1}-\beta_{k_0}}\Phi_{\kappa_2},
\label{sposition}
\end{equation} 
up to a normalized constant. 
The property of such magnonic mode is contained in the Bloch basis $\Phi_{\kappa_1}$ and $\Phi_{\kappa_2}$. When $|\beta_{\kappa_{1,2}}|>1$ ($|\beta_{\kappa_{1,2}}|<1$), the amplitude of the magnonic mode increases with the increase of the site number, leading to the localization at the right (left) boundary.

By the numerical diagonalization of Eq.~\eqref{heff} with the array number $N=51$, the different phonon attenuation, and the different magnon-phonon coupling chiralities, we obtain the frequency spectra $\omega_{\kappa}$ under the OBC. Substituting  $\omega_{\kappa}$ into Eq.~\eqref{energy}, we solve the allowed Bloch wavevector $\kappa$ or $\beta_\kappa=e^{i\kappa d}$. In principle we can also solve the frequency spectra and Bloch wavevector by finding the solutions of Eqs.~(\ref{condition_1}) and (\ref{condition_2}), but this does not provide new solutions.

The equations can also be solved in the PBC, and the solutions are harnessed for addressing the topological properties in the absence of the edge. Correspondingly, the Bloch wavevector is real and ranges from $-\pi/d$ to $\pi/d$ by the translation symmetry. The corresponding frequency spectra is obtained by substituting such real Bloch wavevectors into Eq.~\eqref{energy}. The non-Hermitian skin effect can then be  explained by the so-called winding number that is defined in the PBC as \cite{toposkin,windingnumber2,mutiplewinding}
\begin{equation}
{\cal W}(\omega)=\frac{1}{2\pi}\text{\ensuremath{\oint_{{\rm BZ}}\frac{d}{d\beta_{\kappa}}\text{arg}\left[\hbar\omega_{\kappa}-\hbar\omega\right]d\beta_{\kappa}}},
\label{winding_number}
\end{equation}
where $\hbar \omega$ is the reference energy. ${\cal W}(\omega)$ counts how many times the energy spectra encloses the reference energy when $\beta_{\kappa}$ evolves along the unit circle.

\begin{widetext}
\begin{center}
\begin{figure}[htp]
\centering
\includegraphics[width=13.8cm]{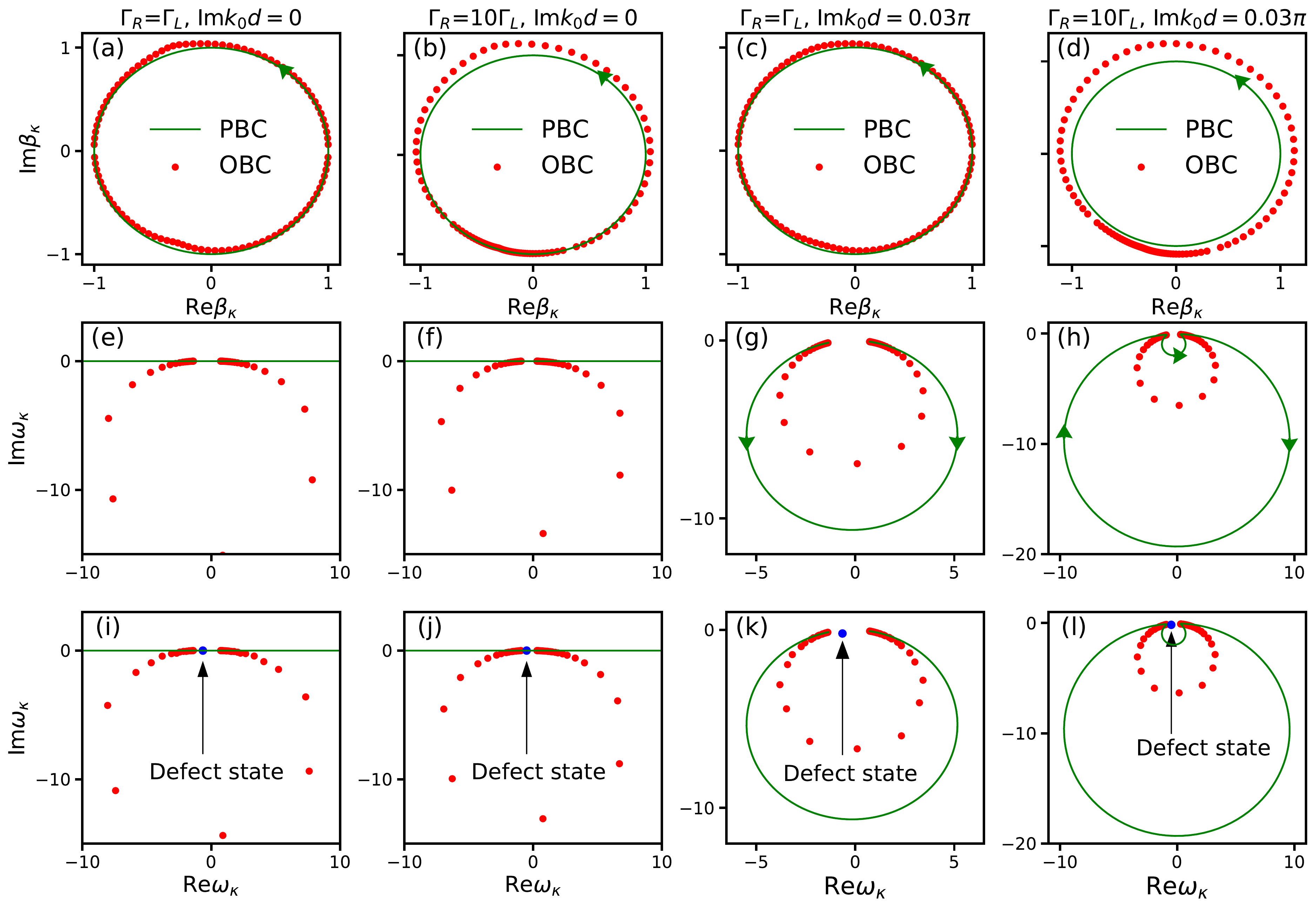}
\caption{The Bloch wavevector $\beta_{\kappa}$, the frequency spectra $\omega_{\kappa}$ [scaled by $(\Gamma_R+\Gamma_L)/2$], and the frequency winding for the magnonic collective modes in the one-dimensional array. The parameters used for the calculation are addressed at the top of the figure. The Bloch wavevector and the frequency spectra are calculated under the OBC (red dots) and the PBC (green line), respectively. (a) to (d) compare the
Bloch wavevector $\beta_{\kappa}$ under the OBC (the red dots) and the PBC (the green unit circle) with different conditions. (e) to (f) address the topological origin of the skin effect by the frequency winding represented by the green curves. (i) to (l) spot the frequency of the defect state in the frequency spectra.}
\label{fig6}
\end{figure}
\end{center}
\end{widetext}

In Fig.~\ref{fig6}(a) to (d), we systematically compare the Bloch wavevector $\beta_\kappa$ under the OBC and the PBC  with different conditions. Under the PBC, the distribution of the Bloch wavevector always lies on a unit circle, with the arrows indicating an evolution of $\kappa$ on the unit circle. Without chirality and phonon attenuation, $\beta_{\kappa}$ is almost overlapped with the unit circle with small exceptions for those superradiant states in Fig.~\ref{fig6}(a), indicating the absence of the non-Hermitian skin effect. Such features and conclusion are not changed  when we only change the chirality or the phonon attenuation alone, as shown in Fig.~\ref{fig6} (b) and (c). However, when we include both the chirality and phonon attenuation, all $\beta_{\kappa}$ deviate from the unit circle, implying the ocurrence of the non-Hermitian skin effect from Eq.~(\ref{sposition}).

We can also trace the topological origin of the above features by the winding number [Eq.~\eqref{winding_number}], as shown in Fig.~\ref{fig6}(e) to (h) by the green curves, which address the evolution of the frequency spectra under the PBC. We find that the winding number is indeed zero in Fig.~\ref{fig6}(e) to (g). However, in the presence of both the chirality and phonon attenuation the winding number is $-1$, as shown in Fig.~\ref{fig6}(h), which leads to the emergence of the non-Hermitian skin effect \cite{WN_and_SE,Topoorigin}. 

When considering a point defect in the middle of the array, the frequency spectra is shown in Fig.~\ref{fig6}(i) to (l). We also plot the frequency spectra under PBC without a point defect by the green curves for the eye guidance. We find in Fig.~\ref{fig6}(i)-\ref{fig6}(l) that all the defect states with different conditions are well isolated and are not enclosed by the frequency windings, such that these defect states cannot be skewed to the boundary and are inertial to the non-Hermitian topology.

\end{appendix}


\end{document}